\documentclass[dvipsnames, sigconf, screen, nonacm]{acmart}

\usepackage{graphicx}
\graphicspath{{./figures/}{./results}}
\DeclareGraphicsExtensions{.pdf}
\usepackage{caption}
\usepackage{subcaption}
\usepackage{tabularray}
\usepackage{tikz}
\newcommand*\circled[1]{\tikz[baseline=(char.base)]{
            \node[shape=circle,draw,inner sep=1pt] (char) {\small {#1}};}}

\begin{document}

\newcommand{\amund}[1]{\textcolor{blue}{Amund: #1}}
\newcommand{\rebuttal}[1]{\textcolor{cyan}{#1}}

\title{\name{}: Realizing Effective Microarchitectures for In-Core Secure Speculation Schemes}

\renewcommand{\sectionautorefname}{Section}
\renewcommand{\subsectionautorefname}{Section}
\renewcommand{\subsubsectionautorefname}{Section}

\newcommand{\name}{ShadowBinding}
\pagenumbering{gobble}

\author{Amund Bergland Kvalsvik}
\email{Amund.Kvalsvik@NTNU.no}
\orcid{ 0009-0007-7112-865X}
\affiliation{%
  \institution{Norwegian University of Science and Technology}
  \city{Trondheim}
  \country{Norway}
}
\author{Magnus Sj\"alander}
\email{Magnus.Sjalander@NTNU.no}
\orcid{0000-0003-4232-6976}
\affiliation{%
  \institution{Norwegian University of Science and Technology}
  \city{Trondheim}
  \country{Norway}
}

\begin{abstract}


Secure speculation schemes have shown great promise in the war against speculative side-channel attacks, and will be a key building block for developing secure, high-performance architectures moving forward. 
As the field matures, the need for rigorous microarchitectures, 
and corresponding performance and cost analysis, 
become critical for evaluating secure schemes and for enabling their future adoption. 


In ShadowBinding, we present effective microarchitectures for two state-of-the-art secure schemes, uncovering and mitigating fundamental microarchitectural limitations within the analyzed schemes, and provide important design characteristics. 
We uncover that Speculative Taint Tracking's (STT's) rename-based taint computation must be completed in a single cycle, creating an expensive dependency chain which greatly limits performance for wider processor cores. 
We also introduce a novel michroarchitectural approach for STT,
named STT-Issue, 
which, by delaying the taint computation to the issue stage, eliminates the dependency chain, achieving better instructions per cycle (IPC), timing, area, and performance results.


Through a comprehensive evaluation of our STT and Non-Spec\-ulative Data Access (NDA) microarchitectural designs on the RISC-V Berkeley Out-of-Order Machine, we find that the IPC impact of in-core secure schemes is higher than previously estimated, close to 20\% for the highest performance core. 
With insights into timing from our RTL evaluation, the performance loss, created by the combined impact of IPC and timing, becomes even greater, at 35\%, 27\%, and 22\% for STT-Rename, STT-Issue, and NDA, respectively.
If these trends were to hold for leading processor core designs, the performance impact would be well over 30\%, even for the best-performing scheme.

Through these findings, research sentiments of Spectre being solvable by in-core secure schemes at low performance costs are challenged.
\name{} serves as a call to arms for more in-depth evaluation of secure speculation schemes, and further work into in-core methods and optimizations, which can help mitigate the high performance cost of current state-of-the-art schemes, without requiring extensive and expensive modifications.
\end{abstract}

\maketitle

\section{Introduction}

The Spectre attacks~\cite{spectre:SP2019} acted as a wake-up call for the architectural community, placing security front-and-center for microarchitectural design.
High-performance processors are dependent on a series of complex microarchitectural optimizations for high performance, including extensive speculation. 
Speculation has introduced a slew of speculative side-channel attacks, which have grown in prominence~\cite{pacman:ISCA2022,dead-uops:ISCA2021}, scope~\cite{netspectre:ESORICS2019,SGXPECTRE:EUROS,tobah:SP2022}, and number~\cite{smotherspectre:CCS2019,ret2spec:CCS2018,Fustos:ARXIV2020, Kiriansky:ARXIV2018,Behnia:ASPLOS2021,spook:SP2022,Li:HPCA2022-unxpec,zombiload:CCS2019} since the release of Spectre~\cite{spectre:SP2019} and Meltdown~\cite{meltdown:SECURITY2018} in 2018.
A multitude of software-based mitigation strategies have been developed, including improvements in compilation~\cite{clearing-shadows:PACT2020}, libraries~\cite{SpectreMSVC,retpoline}, operating systems~\cite{dawg:MICRO2018}, and even microcode patches~\cite{swivel:SECURITY2021, side-channel-mitigations-intel:WEB}.
However, as the attack vector inherently exists in the microarchitecture, microarchitectural mitigation strategies are critical for creating robust and low-cost defenses against these attacks~\cite{Mcilroy:ARXIV2019}, and many secure speculation schemes have been developed over the past few years~\cite{nda:MICRO2019, stt:MICRO2019, dom:ISCA2019, dom:TCOMP2020, ghost:CF2019, ghostminion:MICRO2021, invarspec:MICRO2020, recon:MICRO2023, doppelganger-loads:ISCA2023, muontrap:ISCA2020, cleanupspec:MICRO2019, dolma:SECURITY2021, specshield:PACT2019, specterminator:TACO2022, spt:MICRO2021, safebet:ARXIV2023, invisispec:MICRO2018, hidfix:ICCAD2023}.


Secure schemes come with trade-offs regarding which type of attacks they can block, i.e., how strict they are, and at what performance penalty, with schemes that are both strict and high-performance usually requiring substantial microarchitectural modifications.
Some schemes even employ modifications beyond the core itself, such as the memory hierarchy~\cite{sdo:ISCA2020,ghostminion:MICRO2021,invisispec:MICRO2018} or software~\cite{clearing-shadows:PACT2020,speccfi:SP2019}, to improve performance.
Particularly interesting are in-core secure schemes, which aim to block speculative side-channel attacks without modifying the memory hierarchy or requiring software changes, due to the cost and complexity of such changes. 


\name{}, named for trying to bind the impact of speculation shadows, presents realizable microarchitectural designs for two state-of-the-art in-core secure speculation schemes, namely Non-Speculative Data Access (NDA)~\cite{nda:MICRO2019} and Speculative Taint Tracking (STT)~\cite{stt:MICRO2019}. 
Our microarchitectural designs help uncover important limitations and benefits of the different secure speculation schemes, and uncover novel design challenges that were not previously discovered during more abstract evaluation using architectural simulators.

The original description 
of STT~\cite{stt:MICRO2019} suggests using the register renaming mechanism to track taints, but 
our microarchitectural design uncovers that taint tracking is fundamentally different from register renaming, as taint tracking and register renaming resolve same-cycle dependencies differently, see \autoref{sec:stt-rename}.
We present a functional 
microarchitecture, named STT-Rename, to handle this difference, and highlight how rename tainting creates an expensive, unavoidable dependency chain.  
In addition, we propose an alternative microarchitecture that delays taint tracking until the issue stage, see \autoref{sec:stt-issue}. 
This new 
microarchitecture, named STT-Issue, achieves better results in instructions per cycle (IPC), timing, area, and performance.

For NDA, we show how its limited architectural complexity results in an effective microarchitecture, due to the ease of mapping its design to 
effective hardware. 
Though this microarchitecture performs worse in terms of IPC, it achieves superior results in timing, area, and performance when compared to STT due to its much lower design complexity.



Previous research has primarily evaluated secure schemes using architectural simulators, which, though useful for validating a core premise, often obfuscate microarchitectural design, and provide limited insight into timing, area, and power.
As the field matures, and schemes are being considered for adoption by industry, rigorous microarchitectures and analysis of key design characteristics become critical.

We evaluate our microarchitectural designs through RTL implementations on the RISC-V Berkeley Out-of-Order Machine~\cite{SonicBOOM:CARRV2020}.
The IPC evaluation shows that previously estimated IPC reductions between 8.4\% and 10.7\%, as reported in the literature~\cite{stt:MICRO2019, doppelganger-loads:ISCA2023, recon:MICRO2023, nda:MICRO2019}, might be overly optimistic. 
We instead find that the IPC loss is 18.1\% and 15.5\% for 
STT, and 26.4\% for NDA, with the relative IPC impact worsening as absolute IPC increases. 

\begin{figure}[h]
    \centering
    \includegraphics[width=\linewidth]{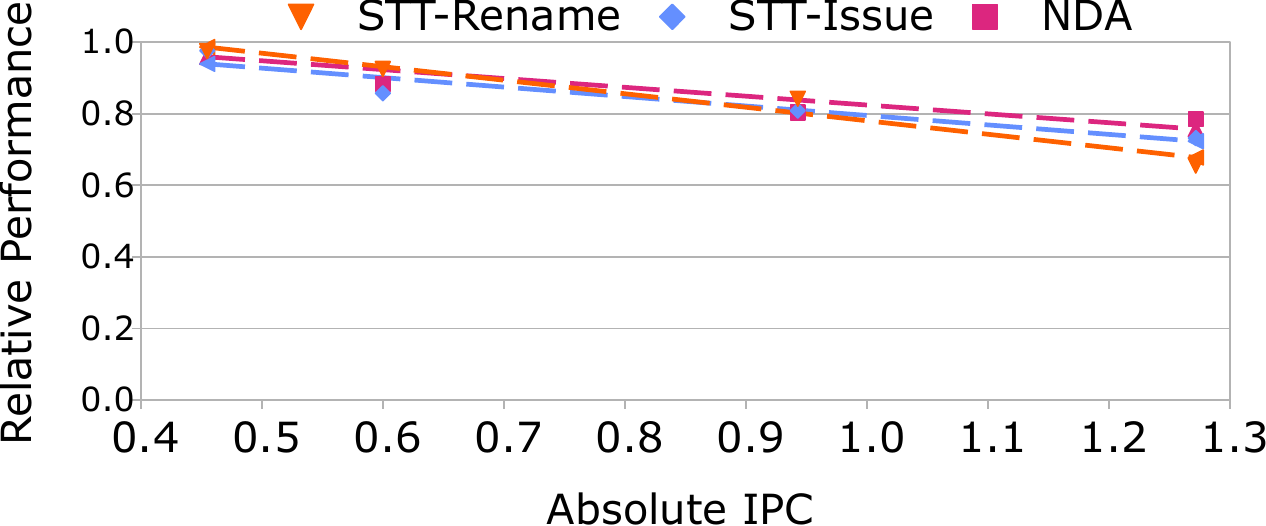}
    \caption{Normalized performance (IPC x Timing) of evaluated secure speculation schemes, with trend line. Data points are placed based on achieved baseline IPC for different configurations.
    }
    \label{fig:scaling-performance}
\end{figure}

When considering performance as the combined effect of IPC and timing, the results are 
even worse.
%
Our evaluation from synthesized RTL designs shows a performance slowdown of 34.5\% and 26.8\% for STT-Rename, and STT-Issue, respectively, and 21.5\% for NDA, when implemented on the highest performance core configuration for the BOOM.
This configuration achieves an average absolute IPC of 1.27 on SPEC CPU2017, see \autoref{fig:scaling-performance}.
This cost is already prohibitively high, but trends indicate that this cost will be even higher for higher performing cores.

Based on a linear extrapolation of our results, a slowdown of more than 40\% is predicted for an Intel Redwood Cove class of core with an IPC of 2.03 for SPEC2017.
Growth is unlikely to be linear, but even with a less pessimistic estimate with only halved growth, the performance slowdown is still a staggering 46.6\% and 37.7\% for STT, and 33.7\% for NDA.
With this less pessimistic trend, NDA outperforms STT by 1.05-1.25$\times$. 

In total, our key contributions are as follows: 
\begin{itemize}
\item 
We uncover that STT’s rename-based taint computation 
must be completed in a single cycle, 
creating an expensive dependency chain,
which scales poorly for wider processor cores.
The design and performance implications of this are shown and evaluated for the first time.
\item We introduce a new microarchitecture for STT, named STT-Issue, which delays tainting to the issue stage, eliminating the expensive dependency chain necessary for STT-Rename, improving IPC, timing, area, and performance results.
\item We elucidate the relationship between secure speculation schemes and overall performance, 
by showing that designs with greater parallelism incur greater relative performance loss, and evaluate this performance impact for two leading in-core state-of-the-art secure speculation schemes.
%
\item 
We show that NDA, perceived as less competitive than STT due to its lower IPC, performs better for our designs, and might be the better solution, as its simpler design translates into better timing results. We augment this insight with detailed performance, timing, area, and power evaluations of the secure speculation schemes. 
\end{itemize}

\name{} serves as a call to arms for renewed efforts into in-core techniques for secure speculation schemes, showing that current techniques are costly, and that 
new in-core 
designs are needed to reduce the costs of secure speculation to an acceptable level. 
Such research needs to precisely consider the microrarchitectural design of existing and new architectural schemes and evaluate the total costs for timing, IPC, and area.
This work is also a first step in answering recent calls for more detailed evaluations as the world is facing a sustainability crisis~\cite{focal-lieven:ASPLOS2024, sustainable-scheduling:ICCAD2024}, as we present results for increased costs in area and power.

\section{Threat Model}
\label{sec:threat-model}

Different secure speculation schemes employ different threat models, i.e., assumptions 
of adversary capabilities and what constitutes a successful attack.
To discuss the different 
schemes, and their evaluation, we give a brief introduction to their 
threat model.

\subsection{Speculative Shadows}
\label{sec:shadows}

To reason about speculative side-channel attacks, it is important to have a robust definition of speculation. 
We use the concept of speculative shadows introduce by Ghost Loads~\cite{ghost:CF2019}, which shows that speculation can occur as a result of control instructions, store-load forwarding, memory consistency, and exceptions, labeling them as C, D, M, and E-shadows, respectively. 
Shadows resolve in-order, and indicate that all following instructions are speculative. 
Our work focuses on C and D-shadows, as these are the most prominent shadows for speculative side-channel attack.
When an instruction has no shadows, it is bound-to-commit, and has reached the visibility point, as it is called by STT~\cite{stt:MICRO2019}.

\subsection{Speculative Taint Tracking (STT)}

STT defends against an adversary that can monitor all covert channels within the system, and can induce speculative execution to access speculative secret data~\cite{stt:MICRO2019}.
Practically, this means that any instruction execution that causes an observable, data-dependent effect, 
is a successful attack if it uses 
speculatively acquired data. 

Secrets are defined as data that the processor would not be able to access during normal execution, i.e., they are the result of transient execution.
As such, entering speculation and speculatively leaking data that resided in registers pre-speculation does not constitute a successful attack, as this data is not considered secret.
STT blocks all data stemming from speculative loads from being used by 
transmitting, i.e., observable, data-dependent instructions, whether there is a direct or indirect data dependency.

\subsection{Non-speculative Data Access (NDA)}

NDA provide several threat models depending on the needs of the user~\cite{nda:MICRO2019}. 
Their two designs, NDA-Strict and NDA-Permissive, have different goals and defend against different threats.
NDA-Strict defends against any leakage of data that would not occur during normal execution.
This model is used for protecting against the leakage of transient data, but also data that resides non-speculatively in registers, and is a stricter scheme than STT. 
NDA-Strict prohibits any propagation of data across the point of speculation, effectively making speculation work as a barrier.

NDA-Permissive defends against the leakage of any data that is speculatively acquired, equivalent to STT. 
This means that the execution of any instruction with a data dependency, directly or indirectly, on a value acquired from speculative execution, is prevented. 
Of note, NDA-Permissive does not require an instruction to be 
observable 
for it to not receive speculative data. 
For this work, we focus on NDA-Permissive, and future references to NDA will be implicitly referencing NDA-Permissive as the evaluated scheme.

\subsection{Combined Threat Model}
Both NDA-Permissive and STT provide equivalent security guarantees for our purposes.
For this work, we consider the two threat models to be equivalent and employ this combined threat model to evaluate security.
As mentioned, we consider speculation stemming from C and D-shadows only. 

\section{Background}
Having established the threat models of the respective secure speculation schemes, we now briefly discuss how STT and NDA each mitigate speculative side-channel attacks. 

\subsection{Speculative Taint Tracking}
\label{sec:background-stt}

As described in the original work~\cite{stt:MICRO2019}, STT employs a form of dynamic information flow tracking (DIFT)~\cite{dift:ASPLOS2004} to track speculative accesses to memory, and their dependency chains to transmitting instructions. 
When a speculative load is renamed, the destination register of the load is marked as tainted, 
with the load set as the taint root. 
Whenever an instruction depends on one or more tainted registers, two things happen:
\begin{itemize}
    \item Firstly, the instruction compares all taint sources it depends on, and selects the youngest of them, setting
    this as its own youngest root of taint (YRoT).
    \item Secondly, the output register of the instruction is marked with the newly computed YRoT, unless the instruction is itself a load, in which case the register is marked with the load as described earlier.
\end{itemize} 

Transmitting instructions that have a YRoT cannot execute until the YRoT source is no longer speculative. 
An instruction is defined as a transmitter if its execution has an observable effect on the system that varies depending on the data in its source operand(s), i.e., its execution is observable and data-dependent.
Non-transmitting instructions can freely execute, regardless of their YRoT state, but they propagate their YRoT information to their output register as described earlier.

As no instructions that are observable can execute while they depend on speculative data, STT ensures that no secret data is broadcast.
Invisible operations, such as common integer operations, are able to execute normally, even while dependent on potentially secret speculative data.
Loads that do not depend on tainted data can execute as normal, helping preserve some memory-level parallelism (MLP) for long-latency loads.

\subsection{Non-speculative Data Access --- Permissive}


NDA-Permissive takes a direct approach to eliminating the propagation of secret data:
Instructions which acquire potentially secret data, i.e., speculative loads, cannot pass this information to any other instructions, until the load is non-speculative, and thereby the data is known to not be a secret. 
To track speculation, NDA propose tracking the oldest point of speculation through the reorder-buffer (ROB), by tracking which instruction types trigger speculation, and under which conditions such speculation is resolved. 

Whenever the oldest point of speculation is resolved, the new oldest head of speculation is computed, and all load instructions older than the new oldest head of speculation are declared safe to propagate.
It is important to note that instructions which do not have a dependency chain on a potential secret are not limited in any way under NDA-Permissive, and can both read source registers, execute, and, unless they are a load, propagate their results as normal.
Speculative loads will delay their writeback and broadcast until they are non-speculative.

Loads that do not depend on speculative data can initiate their memory access as normal, preserving MLP in those cases, and helping hide long-latency loads.

\section{STT Design Considerations}
\label{sec:stt}





In this section, we present two microarchitectures 
for STT, one that performs taint tracking in the rename stages (STT-Rename in \autoref{sec:stt-rename}) and one that delays taint tracking until the issue stage (STT-Issue in \autoref{sec:stt-issue}). 
We also discuss the need for checkpointing when performing taint tracking in the rename stage, as well as some common design considerations, and how the two microarchitectures scale with the width of the core.


\subsection{Taint Tracking During Register Renaming}
\label{sec:stt-rename}

The original STT work~\cite{stt:MICRO2019} does not present a microarchitectural solution for how to propagate taints, but describes taint tracking as similar and compatible with register renaming.
However, as we will show, taint tracking is fundamentally different from register renaming, and when done during register renaming creates an expensive dependency chain that must be performed in a single cycle to ensure correct tainting.

\begin{figure}[t]
    \centering
    \includegraphics[width=1\linewidth]{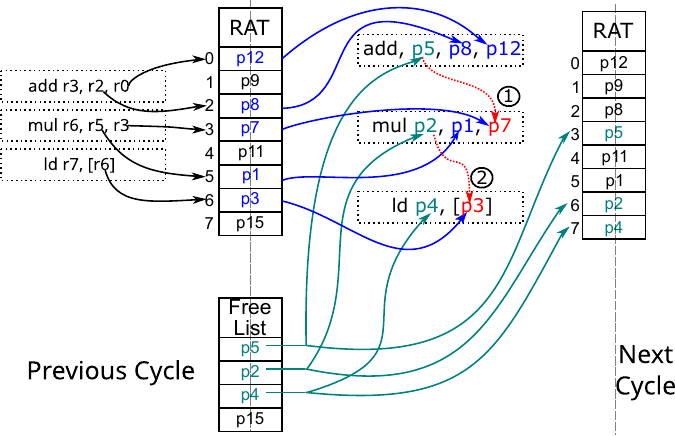}
    \caption{
    A register renaming example with three instructions. 
    Architectural source registers are translated to physical registers through reading the register alias table (RAT), and architectural destination registers are assigned physical registers from the free list. Any errors from same-cycle dependencies, such as \raisebox{.5pt}{\textcircled{\raisebox{-.9pt} {1}}} and \raisebox{.5pt}{\textcircled{\raisebox{-.9pt} {2}}}, are corrected one cycle later. 
    More importantly, the RAT is independently updated with the assigned physical registers from the free list. }
    \label{fig:conventional-rename}
\end{figure}

Register renaming can be performed in two simple steps:
\begin{enumerate}
\item Read all source register from the RAT (\textcolor{blue}{blue}) and for each destination register write a new physical register from the free list to the RAT (\textcolor{teal}{green}), as seen in \autoref{fig:conventional-rename}. 
\item Same-cycle instructions with dependencies, such as exemplified by \raisebox{.5pt}{\textcircled{\raisebox{-.9pt} {1}}} and \raisebox{.5pt}{\textcircled{\raisebox{-.9pt} {2}}}, resolve their errors by replacing the source register read from the RAT with the new physical register for the dependent register. 
\end{enumerate}
Of note is that step~(1) must be performed in a single cycle to ensure that the information in the RAT is up-to-date for the next set of instructions to be renamed in the following cycle.

\begin{figure}
    \centering
    \includegraphics[width=1\linewidth]{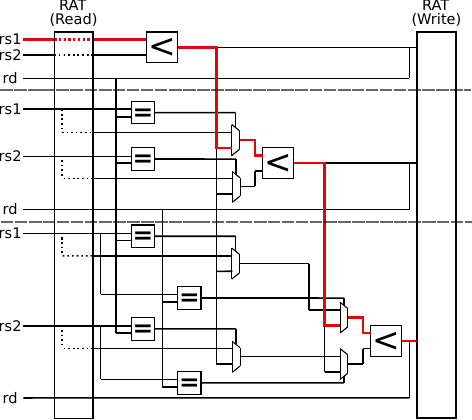}
    \caption{Microarchitecture of YRoT computation for a three instruction-wide rename stage. The critical path is highlighted in red. Dotted lines indicate a translation from index to YRoT. The stippled lines delineate the different instructions. }
    \label{fig:yrot-calc}
\end{figure}

For register renaming, the new register comes from an independent source, namely the free list.
This is fundamentally different for taint tracking, where the new YRoT instead 
depends on the YRoTs of the instructions that the tainting instruction is dependent on, and must be calculated before the new YRoT can be written to the RAT (see \autoref{sec:background-stt}).
Since the YRoT computation of an instruction can depend on the YRoT of 
an older instruction being renamed in the same cycle,
\raisebox{.5pt}{\textcircled{\raisebox{-.9pt} {1}}} and \raisebox{.5pt}{\textcircled{\raisebox{-.9pt} {2}}} in \autoref{fig:conventional-rename}, it is necessary to compute older YRoTs first, causing a chain of YRoT computations and dependency checks, as seen in \autoref{fig:yrot-calc}.

%
The longest potential chain increases linearly with the number of instructions that are renamed in parallel, and all comparisons must be performed in a single cycle to keep the YRoT information in the RAT up-to-date.
The single-cycle requirement introduces a notable timing limitation for wide cores, which negatively impacts overall performance, see \autoref{sec:timing}.



\subsection{Checkpointing During Register Renaming}
\label{sec:checkpoints}

Another challenge with implementing STT, as described in the initial work~\cite{stt:MICRO2019}, is the overlooked cost for supporting branches.
STT, which greatly improved shared 
understanding of implicit transmitters, clarified that branches are a form of transmitter, and should only be resolved when all of its operands are untainted.
In certain scenarios, this means that a younger branch might be resolved before an older branch, if the operands for the younger branch become available and untainted sooner than for the older branch. 
If the branch was incorrectly predicted, it is necessary to restore the architectural state to the point of the younger branch.

This is generally achieved by three mechanisms: A checkpoint of the RAT and free list is stored when a branch is detected in the rename stage; all in-flight instructions younger than the branch are squashed when its mispredict is detected; and the RAT and free list are reset to the stored checkpoint for the mispredicted branch.


As there may still exist live taints in the system, as there may still exist an older source of speculation than the mispredicted branch,
the YRoT information must also be restored. 
This requires that the YRoT information must also be checkpointed whenever a branch is detected in the rename stage.

Additionally, unlike a conventional checkpoint, the YRoT 
state in a checkpoint may be outdated.
Although which YRoT a given instruction depends on will not have changed, the state of the YRoT might have, i.e., the load that is the source of the taint may no longer be speculative.
As such, in addition to restoring from a YRoT checkpoint, it is necessary to also invalidate any entries that are no longer valid due to progress in resolving speculation.
This can be performed by checking the YRoT of each taint and see if it is within the life span for possible YRoT values, i.e., between the youngest load and youngest non-speculative load.

\subsection{Taint Tracking During Instruction Issue} 
\label{sec:stt-issue}

\begin{figure}
    \includegraphics[width=\linewidth]{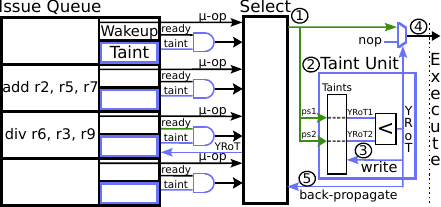}
    \caption{Microarchitecture for STT-Issue. Added structures to support tainting are highlighted in blue. Note that Wakeup and Select are not affected by STT-Issue. Critical path and YRoT depends only on a single instruction unlike for STT-Rename.}
    \label{fig:stt-issue-microarch}
\end{figure}

It is possible to delay tainting and YRoT computation~\cite{secure-by-construction:ICCAD2023}, as tainting only matters once all source operands are ready and the instruction is to be issued.
Such an approach has merit: 
(1)~source operands are more likely to be non-speculative causing fewer tainted destination registers, 
(2)~the total amount of broadcasted YRoT wakeups are reduced, and 
(3)~there are no same-cycle dependency chains as dependent instructions are not issued together. 
We show the 
microarchitecture for such an implementation in \autoref{fig:stt-issue-microarch}.



We describe the STT-Issue microarchiecture in generic terms which are applicable to most scheduling approaches. 
The STT-Issue tainting mechanisms is divided into three main steps:
\vspace{-1ex}
\begin{enumerate}
\item Initially, the YRoT of an instruction is not known and as such the instruction will assume it is safe to issue. 
The wake-up logic operates as normal and emitts a ready signal when all operands have become available.

\item Once the instruction is selected to issue \circled{1}, 
the YRoT for the instruction is computed by a taint unit \circled{2}, which selects the youngest YRoT of all its operands.
If any of the operands are tainted, then the entry corresponding to the physical destination register of the instruction 
is marked as tainted by the calculated YRoT \circled{3}.
If the instruction is tainted and is a transmitter, then it is barred from executing and a no-operation (nop) is issued, which wastes an issue slot for this cycle \circled{4}.

\item Tainted transmitters need to be replayed and woken up once they become non-speculative.
This is achieved by back-propagating the YRoT to the entry in the issue queue \circled{5}. 
A valid YRoT masks the ready signal, preventing the instruction from being selected for issue, until the YRoT is broadcast to be non-speculative, see \autoref{sec:stt-scaling}. 
%
\end{enumerate}
\vspace{-1ex}
High-performance architectures already support speculative issue and replay to improve performance for load dependent instructions~\cite{scheduling-replay:HPCA2004}.
As such, much of the required logic already exists.

A disadvantage of this approach is the increased cost of using physical registers instead of architectural registers.
After rename, all dependencies are tracked through physical registers.
In high-performance processors, the number of physical registers is often an order of magnitude higher than the amount of architectural registers~\cite{chips-redwood-cove}.  
This means that storing taint information and using
comparators for YRoT computation become notably larger.
The total cost of this still remains limited, as we show later in \autoref{tab:area-power}.

Only using physical registers means that there is no need for YRoT checkpoints, see \autoref{sec:checkpoints}.
This is because YRoT values are stored with physical register indexes, which are always live: 
If a register is no longer assigned after a misprediction, it would need to be reassigned, thereby overwriting its previous (stale) YRoT value before it can be used in a YRoT computation.
Similarly, if a register and taint chain is still valid after a misprediction, so too would its values, thereby making YRoT computations correct. 

\subsection{Shared Design Considerations and Scaling }
\label{sec:stt-scaling}

For both microarchitectures, information about the state of taints must be stored, and be accessed for tainting. 
For STT-Rename, this information can be stored together with the RAT, while for STT-Issue it makes more sense to store it in a separate taint unit.

For both STT-Rename and STT-Issue, it is necessary to broadcast whenever a load becomes non-speculative, to inform instructions about the state of their YRoTs.
Whenever the YRoT of an instruction becomes non-speculative, it is safe to execute that instruction, even if it is a transmitter.
This YRoT broadcast functions as an extension of existing broadcast mechanisms, but occurring whenever a load becomes non-speculative.
This broadcast network is expensive, as it requires broadcasting every load that becomes non-speculative to every issue slot, as well as the rename stage or taint unit for STT-Rename and STT-Issue, respectively.  

Though they share some microarchitectural features, 
STT-Issue has several properties that enable it to scale better than STT-Rename.
Key among these is the lack of dependencies between issued instructions, since dependent instructions cannot be issued in the same cycle, which is not the case for register renaming.
The same-cycle dependencies during register renaming create a dependecy chain for STT-Rename, which is not present for STT-Issue. 

%
%
%

As we highlight in \autoref{sec:timing}, STT-Issue pays a higher flat cost in terms of timing, achieving worse timing results than STT-Rename for smaller designs, but shows better 
scaling than STT-Rename due to the lack of same-cycle dependencies.
For smaller designs, STT-Rename may offer higher performance, as we show in \autoref{sec:performance}. 

\section{NDA Design Considerations}
\label{sec:nda}


In this section, 
we present the microarchitecture for NDA.
NDA's simple design translates well into a realizable microarchitecture, only requiring limited modifications to broadcast and otherwise extending mechanisms that already exist within high-performance processor core designs. 
NDA's simplicity results in an insignificant impact on timing, as detailed in \autoref{sec:timing}.

\begin{figure}[t]
    \begin{subfigure}[t]{\linewidth}
    \centering\includegraphics[width=0.75\linewidth]{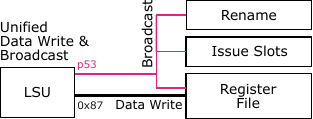}
    \caption{Baseline data write \& broadcast}
    \label{fig:conv_broadcast}
    \end{subfigure}
    \begin{subfigure}[t]{\linewidth}
    \centering\includegraphics[width=0.75\linewidth]{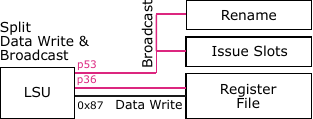}
    \caption{NDA modified data write \& broadcast.}
    \label{fig:nda_broadcast}
    \end{subfigure}
    \caption{Impact of NDA on broadcast and writeback. Note that NDA requires data write be able to select a different physical register than the broadcast. }
\end{figure}

\subsection{Delayed Broadcast}
\label{sec:delayed-broadcast}

NDA requires that the data from a load is not propagated, until the load is non-speculative.
A mechanism is therefore needed that delays broadcasts until speculation is resolved.
Conventionally, when a load request completes, the data is written to the register file, and the register is broadcast as ready.
As loads are often on the critical path, such broadcasts happen immediately upon load completion. 
On load completion, both the broadcast and data writeback refer to the same register, enabling the use of a shared bus, as shown in pink in \autoref{fig:conv_broadcast}.

For NDA, the data writeback is decoupled from the broadcast.
If the load is speculative at the time of completion, then the data is written to the register file, but the broadcast must be delayed until the load becomes non-speculative.
To avoid halving the effective throughput of load operations, the LSU must handle writebacks and broadcasts independently, as shown in \autoref{fig:nda_broadcast}.
This decoupling limits the possibility of sharing a bus for the physical register index and requires a slightly more complex LSU that can perform a writeback in parallel with a broadcast for two separate loads.

We describe mechanisms for tracking speculation for both STT and NDA in \autoref{sec:shadow-tracking}.
Whenever the visibility point increments past a load, a broadcast should be initiated, and several such loads can become non-speculative in the same cycle.
However, the number of parallel broadcasts is limited to the core memory width.

For the purposes of this work, NDA does not support speculative scheduling of dependent instructions by predicting an L1 hit. 
As we show in \autoref{sec:timing}, removing this logic for NDA, which is unlikely to be able to benefit from it regardless, improves the timing of NDA.  

We note that the delayed broadcast mechanism of NDA hinges on whether an instruction is speculative, 
and does not in any way depend on the data that the instruction loads. 
This means that the delayed broadcast mechanism does not introduce any new leakage, as there is no data-dependent behavior for speculative data.

\section{Tracking Speculation}
\label{sec:shadow-tracking}

Both NDA and STT require a way of detecting whether a given load instruction is speculative or not. 
STT needs to track speculation to know when to taint and untaint registers, while NDA needs to track speculation to delay propagation of speculative loads and initiate broadcasts. 

For our work, we only focus on speculation stemming from store-to-load forwarding prediction, and from control speculation, i.e., D and C-shadows~\cite{ghost:CF2019, dom:ISCA2019}, as discussed in \autoref{sec:shadows}. 

High-performance processors already have methods to track C-shadows, which enable them to quickly squash any in-flight instructions that are mispredicted.
Essentially, such a mechanism tracks whether an instruction is dependent on a given branch. 
Regardless of the specifics of the chosen mechanism, we can use such mechanisms to track whether a load is speculative from a branch, by checking whether it could potentially be squashed by the branch tracking mechanism. 

Similarly, because of the risk of address aliasing, load-store units (LSUs) need a method to track D-shadows for memory instructions that might alias.
If aliasing is not detected and checked correctly, stale data could be read in the case of store-to-load forwarding. 
The exact implementation to check for forwarding and forwarding errors will vary based on the microarchitecture, but in all systems, loads which read stale data must flush all following instructions, and restart execution with the correct data.

A simple solution is to check all younger loads for potential address matches whenever a store is committed, and in case of a match, check whether store-to-load forwarding occurred. 
This check can be performed at the earliest when the address of a store is generated. 
For a load, when all stores older than the load have performed such a check, it is known whether a store-to-load forwarding error has occurred, and as such the load is either no longer speculative, or marked as having a forwarding error.

C and D-shadows are most prolific in attacks, and using them as the basis for a secure speculations scheme provides defenses against Speculative Store Bypass~\cite{ssb-intel} and Spectre, but does not protect against the full Futuristic model, as defined in InvisiSpec~\cite{invisispec:MICRO2018}.
Protecting against the Futuristic model would require tracking more speculation points, which can be accomplished by adding M and E-shadows to the speculation tracking system.
An efficient method for tracking shadows is described by Sakalis et al.~\cite{dom:ISCA2019}.

\section{Methodology}
\label{sec:methodology}

To evaluate the overall cost of secure speculation schemes, we implement STT and NDA on the RISC-V BOOM core~\cite{SonicBOOM:CARRV2020}, and evaluate STT with rename tainting, STT with issue slot tainting, and NDA under equal conditions.
We synthesize STT, NDA, and an unsecure baseline for four different BOOM configurations, using AMD Vitis v2022.2 targeting an U250 Alveo FPGA to evaluate timing, area, and power.
Using FireMarshal~\cite{firemarshal:ISPASS2021}, we create Linux images (Ubuntu 18.04) with the SPEC2017 CPU benchmark suite~\cite{SPEC-CPU2017} compiled for RISC-V with GCC 10.1.0. 
Using Firesim~\cite{firesim:ISCA2018}, we boot Linux on the synthesized designs and run the full SPEC2017 CPU benchmark suite, as a representative selection of single-threaded workloads.
Verifying that the implemented schemes mitigate Spectre v1 was done through the use of the BOOM-attacks~\cite{boom-attacks:GITHUB2019}, and intended functionality was verified through waveform analysis.

We collect results by executing each benchmark from the SPEC2017 suite for 100 billion cycles,
examining a large instruction window by taking advantage of the high execution speed of FPGAs compared to simulators, giving us a low margin of error. 

Analyzing hardware performance is normally limited to built-in hardware performance counters, but by using TraceDoctor~\cite{tracedoctor:ISSWC2023} we extract key performance indicators such as committed instructions, latencies, stalls, and their causes. 
With TraceDoctor, we uncover behaviors such as those discussed in \autoref{sec:partial-issues}.

\begin{table}[t]
\caption{The four BOOM configurations, with their key characteristics, and a comparison to the latest Intel processor Redwood Cove, showing average absolute IPC for SPEC2017. \\ \footnotesize $^2$ Redwood Cove supports 3 loads + 2 stores in a cycle. }
\centering
\label{tab:boom-configs}
\begin{tabular}{|c|c|c|c|c|c|}
\hline
             & Small & Medium & Large & Mega & Intel \\ \hline
Core Width   & 1     & 2      & 3     & 4    & 6            \\ \hline
Memory Ports & 1     & 1      & 1     & 2    & 3+2\footnote{Redwood cove has 3 ports for loads and 2 ports for stores that can be used at the same time, while ports are shared in BOOM.}          \\ \hline
ROB Entries  & 32    & 64     & 96    & 128  & 512          \\ \hline
SPEC2017 IPC & 0.46  & 0.6    & 0.943   & 1.27 & 2.03~\cite{spec2017-redwood}       \\ \hline
\end{tabular}
\end{table}

\begin{table}[t]
    \caption{Additional configuration details for gem5.}
    \centering
    \begin{tabular}{|c|c|}
    \hline
         Parameter & Value \\ \hline
         Branch Predictor & MultiperspectivePerceptronTAGE64KB \\ \hline
         L1D Prefetcher &  Stride Prefetcher \\ \hline
         L2 Prefetcher & Stride Prefetcher \\ \hline
    \end{tabular}
    \label{tab:gem5-config}
\end{table}

\begin{figure*}[t]
    \centering
    \includegraphics[width=\linewidth,trim={0cm 0 0cm 0}]{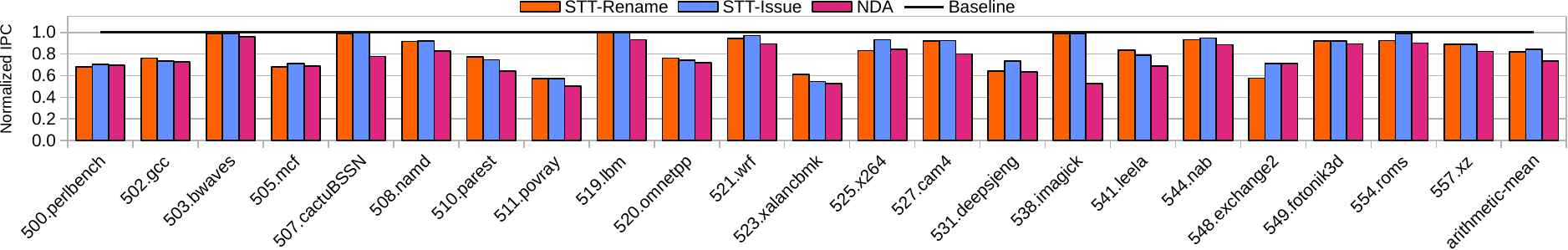}
    \caption{IPC normalized to baseline for different secure speculation schemes for mega size config. }
    \label{fig:ipc-normalized}
\end{figure*}

We also evaluate NDA and STT-Rename on gem5, using the provided configurations in the original works as a guideline. 
We run SPEC2017 in full system simulation mode, using SimPoints gathered from the first 100 billion instructions of each benchmark, with up to five SimPoints for each benchmark, a warmup period of 50 million and an execution period of 50 million instructions. 
Though we use the provided configuration to the best of our ability, information about choice of prefetcher and branch predictor are not provided in the original papers. 
\autoref{tab:gem5-config} shows our configuration for those parameters, which are the same for our NDA and STT evaluation.
We were not able to evaluate the following benchmarks from SPEC2017 using gem5: \texttt{namd, parest, povray}, so for comparisons between BOOM and gem5, we are not including any results for those benchmarks. 

We evaluate the following schemes:
\begin{itemize}
    \item \textbf{Unsafe Baseline:} The unsafe baseline is the unmodified BOOM core which is 
    not protected against Spectre attacks.
    \item \textbf{STT -- Rename:} Speculative Taint Tracking, with taint computation and propagation occurring during the register rename stage, using architectural registers, see \autoref{sec:stt-rename}.
    \item \textbf{STT -- Issue:} Speculative Taint tracking, with taint computation and propagation occurring during the issue stage, using physical registers, see \autoref{sec:stt-issue}.
    \item \textbf{NDA:} Non-speculative Data Access, with a split broadcast and data bus, that broadcasts non-speculative loads, and holds speculative data in registers until they become non-speculative, see \autoref{sec:nda}.
\end{itemize}

We evaluate the secure schemes on four BOOM configurations.
\autoref{tab:boom-configs} shows their key characteristics and the absolute IPC that the unsafe baseline achieves on SPEC2017.
We present the key characteristics for a sense of scale, full configuration details are available on GitHub~\cite{boom:GITHUB}.
We also include the characteristics of an Intel Redwood Cove as a representative example of a recent high-performance core.
By default, we present results for the Mega BOOM configuration, unless another configuration is explicitly mentioned, as it has the highest performance.
In some sections, we explicitly refer to the configurations from \autoref{tab:boom-configs}, and we refer to higher performance configurations with more parallelism as wider.

\section{Results}
\label{sec:results}

Our results highlight key characteristics of the evaluated secure speculation schemes, such as IPC for four different BOOM configurations, and impact on timing, area, power, and total runtime.
We also show comparable results between the BOOM implementations 
and 
equivalent implementations on gem5.

\subsection{Overall Instruction Per Cycle (IPC) Loss}
\label{sec:ipc-results}

\autoref{fig:ipc-normalized} shows the IPC for each of the implemented schemes, normalized to the IPC of the unsafe baseline. 
The results clearly show that the IPC loss of the secure speculation schemes varies greatly depending on the workload.
Some workloads, such as \texttt{503.bwaves}, have insignificant IPC loss, regardless of the chosen scheme.
Other workloads, such as \texttt{538.imagick}, show that versions of STT are close to baseline performance, while NDA suffers a massive slowdown, with nearly half the IPC of baseline.

If a benchmark is compute-bound instead of memory-bound, the IPC loss for STT is limited, as most computations are not transmitters, and can execute irrespective of the taint status of their data dependencies.
For NDA, the IPC loss can still be significant, as the delayed load broadcast means that no dependent computations can be completed, even if the computations themselves are invisible to an attacker.
Benchmarks such as \texttt{507.cactuBSSN} and  \texttt{538.imagick} highlight this.

\begin{figure*}[t]
\begin{subfigure}{1.0\textwidth}
    \includegraphics[width=\linewidth]{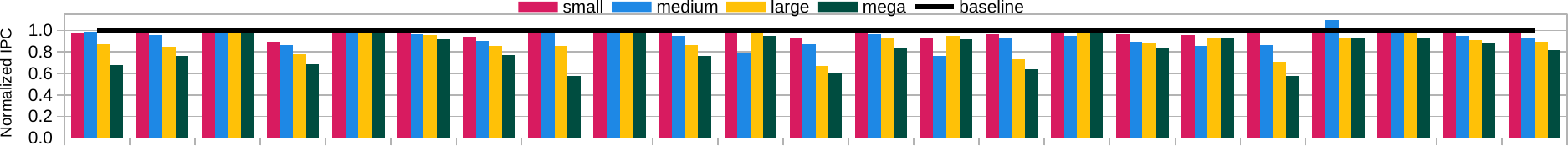}
    \caption{STT-Rename}
    \label{fig:ipc-stt-rename}
    \vspace{2mm}
\end{subfigure}
\begin{subfigure}{1.0\textwidth}
    \includegraphics[width=\linewidth]{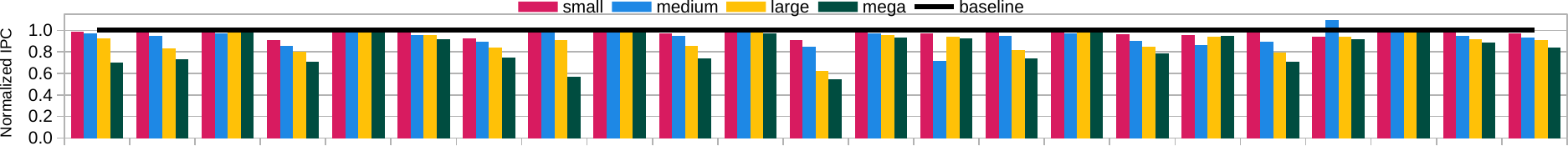}
    \caption{STT-Issue}
    \label{fig:ipc-stt-issue}
    \vspace{2mm}
\end{subfigure}
\begin{subfigure}{1.0\textwidth}
    \includegraphics[width=\linewidth]{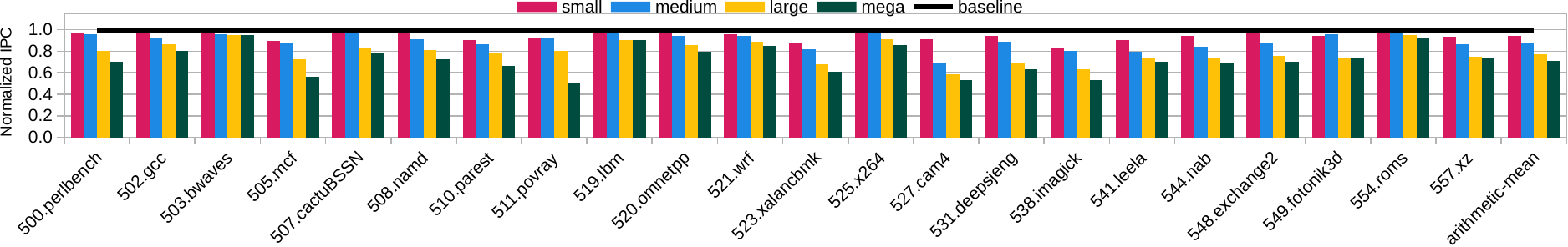}
    \caption{NDA}
    \label{fig:ipc-nda}
\end{subfigure}
\vspace{-5mm}
\caption{Normalized IPC for the four core configurations for (a) STT-Rename, (b) STT-Issue, and (c) NDA.}
\label{fig:size-ipc}
\end{figure*}

\begin{figure}[t]
    \centering
    \includegraphics[width=\linewidth]{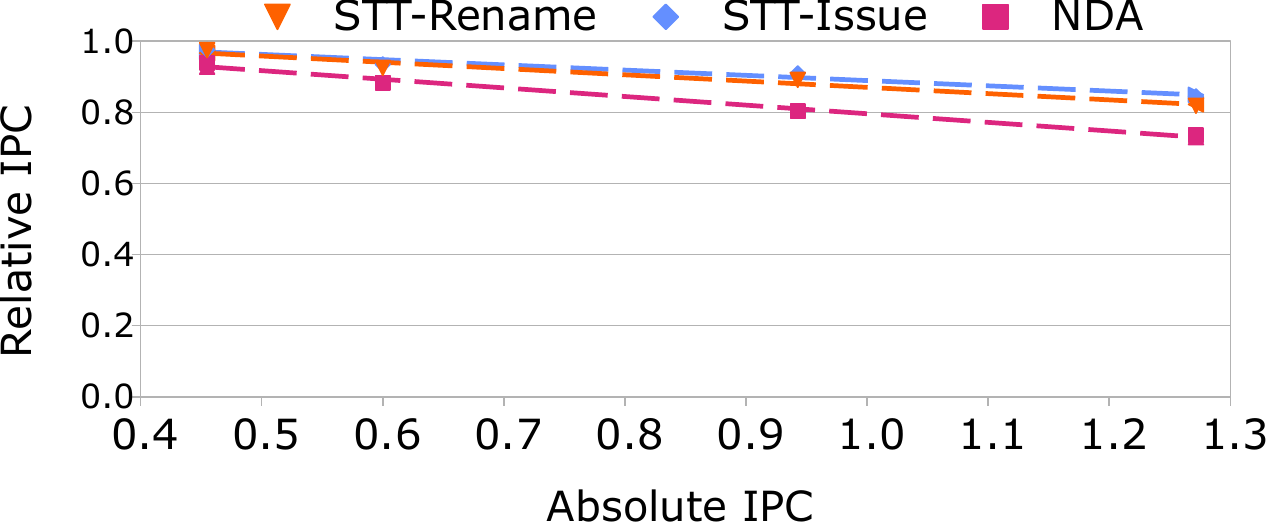}
    \caption{IPC Loss across wider designs, normalized to baseline, with trend line}.
    \label{fig:trend-ipc}
    \vspace{-1mm}
\end{figure}

\texttt{548.exchange2} displays an unexpected behavior, in that NDA achieves higher IPC than both versions of STT, NDA achieving an absolute IPC of 1.77 compared to STT-Rename's 1.44.
Intuitively, one would expect STT to always have equal or better IPC than  NDA, due to being able to execute all the same loads and more load-dependent non-transmitter instructions.
During execution of the benchmark, STT-Rename had 1\,350 as many store-to-load forwarding errors than NDA, losing many cycles handling these errors. 
This is due to STT-Rename preventing store addresses from becoming visible in the LSU, despite it being safe to do so.
Further details are provided in \autoref{sec:partial-issues}.

To calculate average IPC for SPEC2017, we calculate the arithmetic mean of cycles and instructions separately, and calculate the IPC from these averages~\cite{Eeckhout-evaluation-methods:SYN_LECTURE2010}.
The mean shows that STT-Rename, STT-Issue, and NDA achieve on average 81.9\%, 84.5\%, and 73.6\% of baseline IPC, respectively, which is a significant IPC impact. 
The impact is considerably lower for both STT implementations than NDA, and there is a noticeable difference between STT-Rename and STT-Issue. 
Potentially surprising, STT-Issue generally outperforms STT-Rename, due to being able to take advantage of more aggressive scheduling, and avoiding limitations with partial issues.
We explore this in more depth in \autoref{rename-vs-issue} and \autoref{sec:partial-issues}.

\subsection{Correlating IPC Loss With Core Width}

We also evaluate the IPC loss for each of the four BOOM configurations to analyze how the secure speculation schemes scales to wider and higher performance designs.
\autoref{fig:size-ipc} shows the normalized IPC for each of the schemes,
with \autoref{fig:trend-ipc} showing a linear trend with estimated slowdown for a Redwood Cove class of processor.
We include data for all benchmarks to show that the average normalized IPC gets worse with wider cores, and this the trend is consistent across all benchmarks, except for benchmarks that are not meaningfully affected, such as \texttt{bwaves} and \texttt{roms}.
NDA scales worse in general, which is expected as NDA delays more instructions.

For all schemes, as the core width increases, the absolute IPC improves (\autoref{tab:boom-configs}), and the impact on the IPC by the secure schemes gets worse, meaning that more relative IPC is lost the higher the overall IPC of the core is.
Based on estimated trends, though imperfect as designs are not continuous, the total IPC loss of these techniques might be upward of 20\%, or even higher, for leading-edge processors, \autoref{fig:trend-ipc}.

\subsection{Timing Impact}
\label{sec:timing}

In this section, we discuss the impact the secure speculation schemes have on timing after synthesis, i.e., the highest achievable frequency for a given design with and without a secure speculation scheme integrated.
This provides insight into the complexity and performance impact of a given secure speculation scheme. 

\begin{figure*}[t]
  \small
  \setlength{\tabcolsep}{50em}
  \subfloat[Small BOOM]{
    \begin{tblr}{
    colspec = {lc|c|c},
    cell{2}{2} = {Salmon},
    cell{3}{2} = {Salmon},
    cell{4}{2} = {Salmon},
    cell{5}{2} = {Salmon},
    cell{2}{3} = {Salmon},
    cell{3}{3} = {YellowGreen},
    cell{4}{3} = {Salmon},
    cell{5}{3} = {Salmon},
    cell{2}{4} = {YellowGreen},
    cell{3}{4} = {Gray},
    cell{4}{4} = {YellowGreen},
    cell{5}{4} = {YellowGreen},
    }
                   & 160 & 150 & 140 \\ \hline
        Baseline   &     &     &     \\ \hline
        NDA        &     &     &     \\ \hline
        STT-Issue  &     &     &     \\ \hline
        STT-Rename &     &     &     \\ \hline
    \end{tblr}
  }
  \qquad
  \subfloat[Medium BOOM]{
    \begin{tblr} {
    colspec = {c|c|c},
    cell{2}{1} = {Salmon},
    cell{3}{1} = {Salmon},
    cell{4}{1} = {Salmon},
    cell{5}{1} = {Salmon},
    cell{2}{2} = {YellowGreen},
    cell{3}{2} = {YellowGreen},
    cell{4}{2} = {Salmon},
    cell{5}{2} = {YellowGreen},
    cell{2}{3} = {Gray},
    cell{3}{3} = {Gray},
    cell{4}{3} = {YellowGreen},
    cell{5}{3} = {Gray},
    }
        130 & 120 & 110 \\ \hline
            &     &     \\ \hline
            &     &     \\ \hline
            &     &     \\ \hline
            &     &     \\ \hline
    \end{tblr}
  }
  \qquad
  \subfloat[Large BOOM]{
    \begin{tblr} {
    colspec = {c|c|c|c|c},
    cell{2}{1} = {Salmon},
    cell{3}{1} = {Salmon},
    cell{4}{1} = {Salmon},
    cell{5}{1} = {Salmon},
    cell{2}{2} = {YellowGreen},
    cell{3}{2} = {YellowGreen},
    cell{4}{2} = {Salmon},
    cell{5}{2} = {Salmon},
    cell{2}{3} = {Gray},
    cell{3}{3} = {Gray},
    cell{4}{3} = {Salmon},
    cell{5}{3} = {Salmon},
    cell{2}{4} = {Gray},
    cell{3}{4} = {Gray},
    cell{4}{4} = {Salmon},
    cell{5}{4} = {YellowGreen},
    cell{2}{5} = {Gray},
    cell{3}{5} = {Gray},
    cell{4}{5} = {YellowGreen},
    cell{5}{5} = {Gray}
    }
        100 & 95 & 90 & 85 & 80 \\ \hline
            &    &    &    &    \\ \hline
            &    &    &    &    \\ \hline
            &    &    &    &    \\ \hline
            &    &    &    &    \\ \hline
    \end{tblr}
  }
  \qquad
  \subfloat[Mega BOOM]{
    \begin{tblr} {
    colspec = {c|c|c|c|c|c},
    cell{2}{1} = {Salmon},
    cell{3}{1} = {Salmon},
    cell{4}{1} = {Salmon},
    cell{5}{1} = {Salmon},
    cell{2}{2} = {Salmon},
    cell{3}{2} = {YellowGreen},
    cell{4}{2} = {Salmon},
    cell{5}{2} = {Salmon},
    cell{2}{3} = {YellowGreen},
    cell{3}{3} = {Gray},
    cell{4}{3} = {Salmon},
    cell{5}{3} = {Salmon},
    cell{2}{4} = {Gray},
    cell{3}{4} = {Gray},
    cell{4}{4} = {Salmon},
    cell{5}{4} = {Salmon},
    cell{2}{5} = {Gray},
    cell{3}{5} = {Gray},
    cell{4}{5} = {YellowGreen},
    cell{5}{5} = {Salmon},
    cell{2}{6} = {Gray},
    cell{3}{6} = {Gray},
    cell{4}{6} = {Gray},
    cell{5}{6} = {YellowGreen}
    }
        85  & 80 & 75 & 70 & 65 & 60 \\ \hline
            &    &    &    &    &    \\ \hline
            &    &    &    &    &    \\ \hline
            &    &    &    &    &    \\ \hline
            &    &    &    &    &    \\ \hline
    \end{tblr}
  }

  \caption{Achieved timings (in MHz) during synthesis for four different BOOM configurations, see \autoref{tab:boom-configs}. \textcolor{YellowGreen}{Green}: successfully met timing, \textcolor{Salmon}{red}:~did not meet timing, and \textcolor{Gray}{gray}: better result available.}
  \label{fig:timing}
\end{figure*}

\begin{figure}[t]
    \centering
    \includegraphics[width=\linewidth]{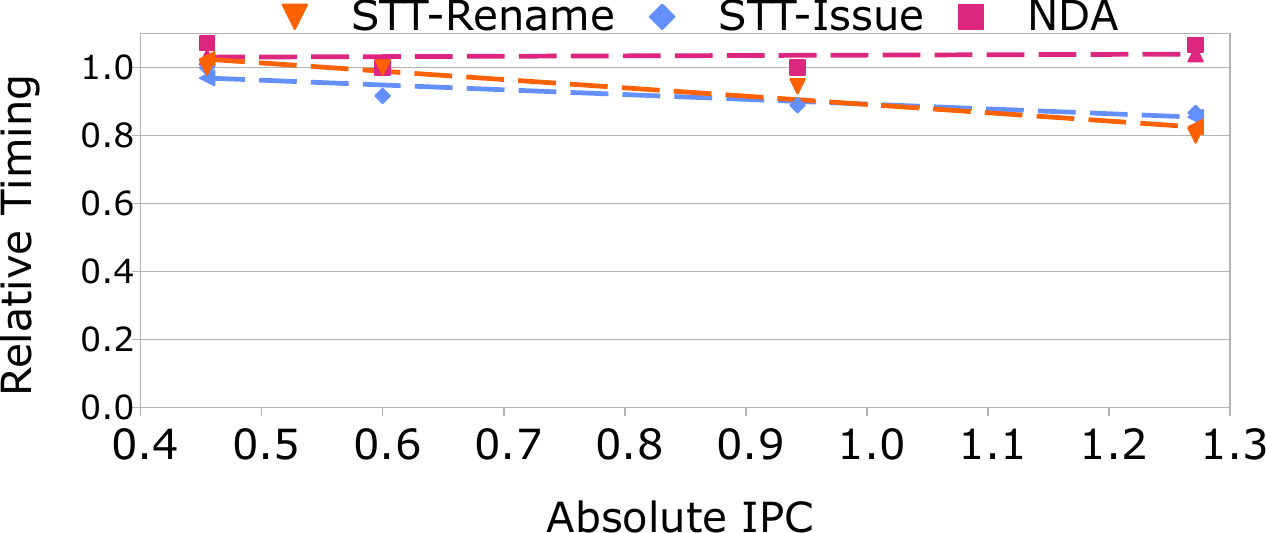}
    \caption{Best timing results for different core configurations, normalized to baseline, with trend line}.
    \label{fig:timing-trend}
\end{figure}

\autoref{fig:timing} shows the highest achieved frequency for each secure speculation scheme for the Small, Medium, Large, and Mega BOOM configurations.
The results indicate that the timing impact of STT increases for wider core designs, while NDA consistently achieves the same, or even higher, frequency as the unsafe baseline. 
This is expected as NDA does not introduce much additional logic, and due to not supporting speculative L1 hit broadcasts, simplifies the core design, achieving better timing for some configurations. 

STT-Rename and STT-Issue both significantly limit timing for the wider configurations.
As explained in \autoref{sec:stt-rename}, STT-Rename scales poorly due to the YRoT computation relying on a chain of dependencies that must be resolved in a single cycle (\autoref{fig:yrot-calc}), resulting in greater timing impact the more instructions are renamed in parallel.
For the four-wide Mega BOOM core, STT-Rename achieves only 80\% frequency compared to the unsafe baseline. 
STT-Issue does not have such a chain of dependencies, as its YRoT computations are independent. 
As tainting is performed a timing-sensitive stage of the 
core, there is still a notable cost in timing. 

\autoref{fig:timing-trend} shows the trend in relative timing compared to the unsafe baseline when the core width increases.
The relative timing of NDA does not change with core-width, while STT-Issue sees a notable impact for the Medium configuration, but with only slight increases for wider designs. 
STT-Rename, on the other hand, has a small timing impact for smaller configurations, but the impact grows for wider cores. 
These trends might not hold for all wider designs, as design constraints vary greatly, but generally wider configurations and designs should experience greater timing limitations for STT-Rename and STT-Issue, with NDA likely to achieve similar timings as baseline.

\subsection{Performance = IPC $\times$ Timing}
\label{sec:performance}

Performance is the combination of IPC and timing, and \autoref{table:performance} shows the performance for each of the secure speculation schemes, normalized against the unsafe baseline.
The trends across core widths was previously shown in \autoref{fig:scaling-performance}, which highlights that the wider a core gets, the higher impact the secure speculation schemes have on overall performance.
This is due to their increasing IPC impact (for all schemes) and timing limitations (for STT), as the core width increase.

\begin{table}[h]
    \centering
    \caption{Tabled data of \autoref{fig:scaling-performance}, normalized performance for configurations, with halved linear estimates for an Intel Redwood Cove class processor.}
    \label{table:performance}
    \begin{tabular}{|l|c|c|c|c|c|}
    \hline
        Scheme     & Small & Medium & Large & Mega & Intel \\ \hline
        STT-Rename & 0.98 & 0.93 & 0.84 & 0.65 & 0.53 \\ \hline
        STT-Issue  & 0.98 & 0.86 & 0.81 & 0.73 & 0.62 \\ \hline
        NDA        & 1.01 & 0.88 & 0.80 & 0.78 & 0.66 \\ \hline
    \end{tabular}
\end{table}

STT has been viewed by the research community as outperforming NDA due to reporting a lower reduction in IPC. 
However, we show that for our RTL based design, that NDA scales better than STT due to its simpler design, which incurs lower penalties in terms of timing for wider cores.
NDA has an overall higher performance, already at the Mega BOOM configuration, with trends indicating a growing gap.
This gap is notable for STT-Rename, but less significant for STT-Issue, due to STT-Issue having better timing and IPC results than STT-Rename.
Even if NDA could achieve only baseline timing, it would still have higher performance than both versions of STT for both Mega BOOM (0.74) and Redwood Cove class processor (0.64).
This challenges previous analysis that used only use IPC.
We delve further into this in \autoref{sec:gem5-discussion}.

\subsection{Area and Power}
\begin{table}[t]
    \centering
    \caption{Area (lookup tables (LUTs) and flip-flops (FFs)), and power results normalized to the baseline for all schemes synthesized at 50 MHz.}
    \begin{tabular}{|c|c|c|c|}
    \hline
       Scheme & LUTs & FFs & Power \\ \hline
       STT-Rename & 1.060 & 1.094 & 1.008 \\ \hline
       STT-Issue & 1.059 & 1.039 & 1.026 \\ \hline
       NDA & 0.980 & 1.027 & 0.936 \\ \hline
    \end{tabular}
    \label{tab:area-power}
\end{table}

\autoref{tab:area-power} shows the area and power increase for each of the secure speculation schemes when synthesized at 50 MHz. 
While both STT-Rename and STT-Issue have a similar increase in lookup tables (LUTs), STT-Rename has a considerably higher increase in flip-flops (FFs), due to its need for checkpoints, see \autoref{sec:checkpoints}.
Both of these increases for STT are significant, while NDA sees a reduction in LUTs, due to simpler logic, but has a slight increase in FFs. 

The power remains unchanged for STT-Rename, increasing slightly for STT-Issue, and decreasing notably for NDA.
This means that in terms of sustainability~\cite{focal-lieven:ASPLOS2024}, NDA has a significant edge in all categories compared to STT-Rename and STT-Issue.

\subsection{gem5}
\label{sec:gem5}

For completeness’s sake, we also implement NDA and STT-Rename on gem5, and run the SPEC2017 suite, comparing benchmarks and overall results with the most comparable BOOM implementations. 

As can be seen from \autoref{table:gem5}, the NDA configuration achieves a relatively low absolute IPC using the given configuration, with an IPC between the Medium and Large configuration.
The overall IPC impact is lower than expected on the gem5 evaluation than the BOOM evaluation, compared to what is expected for the given baseline absolute IPC.

STT achieves a relatively high baseline absolute IPC with its configuration, similar to the absolute IPC achieved by the Mega configuration.
The IPC impact is nearly identical for the BOOM Mega and the gem5 implementation, for STT-Rename.

We discuss the implications of these results in \autoref{sec:gem5-discussion}.

\begin{table}[t]
    \centering
    \caption{IPC for the Medium, Large, and Mega BOOM, and gem5, for STT-Rename, STT-Issue, and NDA. \\ \footnotesize $^3$ using STT gem5 configuration~\cite{stt:MICRO2019}. $^4$ using NDA gem5 configuration~\cite{nda:MICRO2019}. \normalsize}
    \small
    \label{table:gem5}
    \begin{tabular}{|c|c|c|c|c|}
    \hline
        Configuration & Baseline& STT-Rename & STT-Issue & NDA\\
         & Abs. IPC & IPC Loss & IPC Loss & IPC Loss\\ \hline
        BOOM Medium & 0.54 & 7.3\% & 6.4\% & 10.7\% \\ \hline
        BOOM Large & 0.83 & 11.3\% & 10.0\% & 18.6\% \\ \hline
        BOOM Mega  & 1.09 & 17.6\% & 15.8\% & 22.4\% \\ \hline 
        gem5  & 1.12$^3$/0.79$^4$ & 17.2\% & N/A & 13.0\% \\ \hline
    \end{tabular}
    \normalsize
\end{table}

\section{Discussion}
\label{sec:discussion}

In this section, we discuss the feasibility implications for secure schemes, performance phenomenons, the scaling design costs of the schemes, and why our results differ from earlier evaluations, emphasizing our contributions contra those using gem5.

\subsection{STT-Rename vs. STT-Issue}
\label{rename-vs-issue}

For most of the SPEC2017 suite, STT-Issue outperform STT-Rename. 
This might seem counterintuitive, as STT-Issue has the potential of issuing instructions that will be killed in-flight (\autoref{sec:stt-issue}), potentially wasting resources, something that STT-Rename avoids, by tainting instructions and determining their YRoT during rename. 
However, the difference comes down to at what point an instruction gets scheduled to issue and when the YRoT gets calculated.

An instruction is conventionally scheduled for issue as soon as all its operands are ready.
Using STT-Issue, an instruction has not yet determined if it is tainted, and will be scheduled as normal.
Using STT-Rename, on the other hand, tainting has already occurred, and an instruction might therefore be blocked due to being tainted. 
Thus, a potentially tainted instruction can be selected to issue under STT-Issue, and if its YRoT is declared safe in the following cycle, the instruction will execute and complete, as the YRoT will be safe at the time of determining taints.
This enables STT-Issue to, in certain cases, issue instructions one cycle earlier than STT-Rename.
In benchmarks for which STT-Issue outperforms STT-Rename, this and the advantages described in \autoref{sec:partial-issues}, are the main reasons.

Overall, as STT-Issue outperforms STT-Rename in all categories for our evaluation,
STT-Issue is generally preferred instead of STT-Rename, unless there are other factors to consider, such as the issue stage being more expensive to modify.

\subsection{Partial Issues and Store-to-Load Forwarding}
\label{sec:partial-issues}

Store instructions consist of an address and a data element. 
Depending on the specifics of the implementation of a given ISA, a store instruction might be split into an address and a data micro-op, might be a unified micro-op for both that can partially issue whenever one operand is ready, or might be a single micro-op for both that can only issue whenever both operands are ready.
Issuing the address generation early is beneficial, as it makes the address of a store available in the store queue, and thereby enables store-to-load forwarding checks to proceed earlier, lowering the chances of store-to-load forwarding errors. 

In the BOOM core, stores are a single micro-op that can partially issue whenever either the data or address element are ready, with the other half issuing whenever its other operand becomes ready. 
However, with STT, this now becomes more complicated. 
When the YRoT is computed, it uses both operands to find the YRoT, meaning that partial issues for address calculation might be blocked, not because the address operand is not ready, but because the instruction is tainted, and its YRoT is not safe.

In the case of \texttt{exchange2}, a sudoku solver~\cite{exchange2}, which uses a lot of memory operations that span very small memory spaces, this results in the STT implementations having a considerably higher amount of store-to-load forwarding errors, because the store addresses are delayed.
STT-Issue suffers less from this problem, as it does not calculate its YRoT for the entire store if only part of its operands are ready, and can as such issue partial, untainted address generation in most cases.
An optimization could enable STT-Rename to work the same way, by having two taints, one for each operand in the specific case of stores, or by fully splitting the store operation into two micro-ops, one for address and one for data. 
This type of minute details can have significant performance and design implications for certain workloads.

\subsection{Microarchitectural Complexity}

An important consideration for selecting the right secure speculation scheme for an architecture is the design cost such a scheme will create. 
Design cost can be measured through physical results, such as effects on timing and area overhead.
We document such measurements in \autoref{fig:timing} and \autoref{tab:area-power}, which indicate that 
STT-Rename incurs higher costs than 
STT-Issue.

In addition to direct physical design costs, we can evaluate 
microarchitectural complexity.
As mentioned earlier, NDA incurs limited
changes, only modifying aspects of broadcast for loads specifically. 
This is in contrast to STT, which modifies every instruction that passes through rename/issue, to ensure that its DIFT conventions are followed.
Similarly, every instruction that can be delayed by a taint, i.e., all transmitters, need to have a broadcast mechanism for when its YRoT is safe. 
Regardless of the specifics of the underlying core, such modifications will be more complex, more invasive, demand a higher amount of area, and have graver impact on timing compared to NDA. 
However, if timing costs are not incurred, STT-Issue might outperform NDA due to higher IPC. 

\subsection{Trends for Wider Designs}

A key contribution of this work is the investigation into the performance cost of secure speculation schemes by using hardware prototypes, and determining the scalability of the evaluated schemes.
As seen by the performance impact shown in \autoref{fig:scaling-performance}, and the impact on IPC and timing in \autoref{fig:size-ipc} and \autoref{fig:timing-trend}, respectively, performance trends are more complex than simply isolating the impact on IPC for a given configuration.
Our results clearly highlight that many aspects of the evaluated secure speculation schemes 
likely scale poorly with wider core designs, an issue potentially exacerbated even further for modern processors, which commonly boast a core width of six, or more, instructions.

From our analysis, NDA shows better overall performance than both versions of STT, and considerably better than STT-Rename, despite its lower IPC, due to having a much lower impact on timing. 
Similarly, NDA has the lowest impact on area overhead of the three schemes, and is the only scheme with 
reduced power consumption compared to baseline.
From our analysis, when evaluating sustainability, 
NDA is the preferable choice, regardless if the embodied or operational equivalent greenhouse gas footprint is dominating~\cite{sustainability-lieven:CAL2022, focal-lieven:ASPLOS2024}.

STT-Rename shows the worst results in all categories (except IPC compared to NDA), clearly showing that even if STT's higher IPC is desired, the originally proposed 
microarchitecture is not optimal. 
This work highlights the need for future work 
finding more efficient in-core solutions to speculative side-channel attacks.

\vspace{-1mm}
\subsection{Comparing to gem5}
\label{sec:gem5-discussion}

Previously reported performance results for gem5 vary greatly between different implementations and evaluations~\cite{stt:MICRO2019, nda:MICRO2019, doppelganger-loads:ISCA2023, recon:MICRO2023, dolma:SECURITY2021, muontrap:ISCA2020}.
We note that many works use SPEC CPU2006, which is outdated, and results in different IPC trends than with the updated 2017 version.
Even accounting for this, there is a notable gap in previously reported results and ours.
The reasons for this is multifold:

Firstly, gem5 allows users to freely modify key core parameters, such as memory latency. 
Changing these parameters can obscure key details or lead to incorrect conclusions if the chosen configuration is not representative.
For example, earlier works have evaluated STT with 
a single cycle latency for the L1 data cache, which is 3--4 cycles faster than the latest Intel processors~\cite{rf-prefetching:ISCA2022}.
If memory is a notable limitation of a given secure speculation scheme, incorrect memory behavior might give misleading results.

Secondly, simulators such as gem5 are slower than FPGAs.
This makes evaluating long benchmark suites such as SPEC2017 difficult, as running the entire benchmark is not feasible. 
Many evaluations run only small portions of a benchmark, which might not capture all or even a representative selection of benchmark phases, giving inaccurate results.
Ideally, SimPoints based on all or a large part of the benchmark should be used to provide more reliable numbers.

Finally, due to first implementing the schemes using RTL, we were able to create an equivalent microarchitcture in gem5, which achieves a similar IPC to our BOOM implementation.
Hardware prototyping provides a better guidance for microarchitecture,
as the constraints given emulate a real design, and it is therefore harder 
to make infeasible configurations and design choices.


Previously, STT~\cite{stt:MICRO2019} reported a performance overhead of 8.5\%, while NDA~\cite{nda:MICRO2019} reported a performance overhead of 22.3\%.
The reported results for NDA correlate well to our BOOM Mega configuration results, but our results for STT are more pessimistic, showing equivalent performance impact to the smaller BOOM Large configuration. 
We emphasize the need for hardware prototyping as a tool for indicative results and design analysis when performing such evaluations, as relying purely on architectural simulators can give an imprecise understanding of performance.

Critically, conclusions such as STT outperforming NDA due to lower IPC reduction, are incorrect when other factors, such as timing, are considered, which cannot be adequately assessed by current architectural simulators.

\vspace{-1mm}
\section{Related Work}
\label{sec:related-work}

There have been limited investigations into the practical cost of secure speculation schemes using hardware designs. 
Other work has previously implemented STT on the BOOM core, but none have to our knowledge focused specifically on performance evaluations of secure speculation schemes.

Tobias Jauch et al., in their Secure-by-Construction work~\cite{secure-by-construction:ICCAD2023}, implement STT as part of their framework for verifying secure properties.
Their work focused on finding examples of leakage, 
and can efficiently analyze all microarchitectural forms of information leakage. 
Their implementation of STT is based around being comprehensive, and generic, tainting post register-read instead of post-issue, and communicating execution status broadly, which incurs extra area cost.

Other mitigation strategies have employed modifications outside the core to mitigate Spectre, such as InvisiSpec~\cite{invisispec:MICRO2018}, MuonTrap~\cite{muontrap:ISCA2020}, GhostMinion~\cite{ghostminion:MICRO2021}, and SDO~\cite{sdo:ISCA2020}.
As we focus on less complex in-core techniques, these fall outside the scope of this work, but such strategies show promising performance results, though their current evaluation is limited to simulator based approaches.
Modifications to the cache hierarchy are generally more invasive than in-core schemes, though they may be necessary if the performance penalties found with in-core schemes are unavoidable.

\section{Conclusion}

In this work, we have presented several novel insights into the microarchitecutral designs and their implications for state-of-the-art secure speculation schemes.
We have uncovered how STT-Rename's tainting necessitates a chain of dependencies, which must be resolved in a single cycle to keep the taint information correct. 
We present an effective microarchitecture for STT-Rename which addresses the original limitations of tainting during the rename stage, 
as well as an alternative design, 
STT-Issue, 
which eliminates the chain of dependencies by delaying tainting until issue, achieving better performance, timing, area, and power results.
We have shown how NDA's limited complexity easily translates into an effective microarchitecture, with only limited design impact. 

We have provided rigorous microarchitectures and corresponding design characteristics for both NDA and STT. 
Importantly, our results challenge established performance evaluation for secure schemes, highlighting that the real cost of Spectre mitigation might be much higher than previously estimated.
As the field matures, such evaluation become critical for scheme evaluation and industry adoption.
We are, to the best of our knowledge, the first work to present timing, area, and power analysis for secure speculation schemes based on synthesis of microarchitectural RTL designs. 

In total, our evaluation puts a 
spotlight on the importance of detailed evaulation for secure speculation schemes, and invites future research into better solutions for in-core schemes, as current designs incur higher overheads than previously reported. 
Particularly, this motivates further investigation into optimization strategies such as InvarSpec~\cite{invarspec:MICRO2020}, Doppelganger Loads~\cite{doppelganger-loads:ISCA2023}, and ReCon~\cite{recon:MICRO2023}, to help improve the limited IPC of underlying schemes. 

\balance

\bibliographystyle{ACM-Reference-Format}
\bibliography{refs}

\end{document}